\documentclass{appolb}
\usepackage[utf8]{inputenc}
\usepackage{cite}
\usepackage{authblk}
\usepackage{mathtools, cuted}
\usepackage{graphicx}
\usepackage{cancel}
\usepackage{xcolor}
\usepackage{longtable}

\def\ii{\imath 0}
\def\Eq#1{Eq.~(\ref{#1})}
\def\beq{\begin{equation}}
\def\eeq{\end{equation}}
\def\bea{\begin{eqnarray}}
\def\eea{\end{eqnarray}}
\def\nn{\nonumber}
\def\qon#1{q_{#1,0}^{(+)}}

\def\qb{\mathbf{q}}

\def\ket#1{|{#1}\rangle}
\def\bra#1{\langle{#1}|}
\def\id{\boldsymbol I}
\renewcommand{\theenumi}{\roman{enumi}}

\begin{document}
\title{From Causal Representation of Multiloop Scattering Amplitudes to Quantum Computing
\thanks{Presented at Matter To The Deepest 2021 conference.}
}
\author{Selomit Ram\'irez-Uribe
\address{Instituto de F\'isica Corpuscular, Universitat de Val\`encia -- Consejo Superior de Investigaciones Cient\'ificas, Parc Cient\'ific, E-46980 Paterna, Valencia, Spain.}
\address{Facultad de Ciencias de la Tierra y el Espacio, Universidad Aut\'onoma de Sinaloa, Ciudad Universitaria, CP 80000 Culiac\'an, Mexico.}
\address{Facultad de Ciencias F\'isico-Matem\'aticas, Universidad Aut\'onoma de Sinaloa, Ciudad Universitaria, CP 80000 Culiac\'an, Mexico.}
\\
}

\maketitle
\begin{abstract}
An overview of a quantum algorithm application for the identification of causal singular configurations of multiloop Feynman diagrams is presented. 
The quantum algorithm is implemented in two different quantum simulators, the output obtained is directly translated to causal thresholds needed for the causal representation in the loop-tree duality.
\end{abstract}
  
\section{Introduction}

Quantum computing is a natural advantageous framework for problems where the quantum principles of superposition and entanglement can be exploited.
It is currently an approach with great potential in physics~\cite{Feynman:1981tf} to tackle problems that are too demanding for classical computers because they scale exponentially or superpolynomially.

Currently, quantum algorithms are becoming a focus of attention in high-energy physics given the high demands that the field~\cite{Strategy:2019vxc} will face in the coming Run 3 of the CERN's Large Hadron Collider (LHC), the planned phase of high-luminosity~\cite{Gianotti:2002xx}, and the different projects concerning future colliders~\cite{Abada:2019lih,Djouadi:2007ik,Roloff:2018dqu,CEPCStudyGroup:2018ghi}. 
The latest applications in this area consider lattice gauge theories~\cite{Jordan:2011ne,Banuls:2019bmf,Zohar:2015hwa,Byrnes:2005qx}, the speed up of jet clustering algorithms~\cite{Wei:2019rqy,Pires:2021fka,Pires:2020urc}, jet quenching~\cite{Barata:2021yri}, simulation of parton showers~\cite{Bauer:2019qxa,Bauer:2021gup}, determination of parton densities~\cite{Perez-Salinas:2020nem}, heavy-ion collisions~\cite{deJong:2020tvx} and quantum machine learning~\cite{Guan:2020bdl, Wu:2020cye,Trenti:2020ceh}. 

In this paper the problem to be addressed is the determination of the causal thresholds of multiloop Feynman integrals from the identification of all internal configurations that fulfill causal conditions.  
This problem can be targeted by applying a modified version of Grover's quantum algorithm~\cite{Grover:1997fa} for querying multiple solutions over unstructured databases~\cite{Boyer:1996zf}. 

The LTD framework~\cite{Catani:2008xa,Rodrigo:2008fp,Bierenbaum:2010cy,Bierenbaum:2012th,Tomboulis:2017rvd,Runkel:2019yrs,Capatti:2019ypt} opens any loop diagram into a sum of connected trees. 
This methodology has been deeply studied~\cite{Buchta:2014dfa,Buchta:2015xda,Hernandez-Pinto:2015ysa,Jurado:2017xut,Driencourt-Mangin:2019yhu,Aguilera-Verdugo:2019kbz} and many applications have been developed~\cite{Buchta:2015wna,Sborlini:2016gbr,Sborlini:2016hat,Driencourt-Mangin:2017gop,Driencourt-Mangin:2019aix,Capatti:2019edf,Plenter:2019jyj,Prisco:2020kyb,Plenter:2020lop,Runkel:2019zbm}. 
In recent years the LTD has evolved in a significant way~\cite{Verdugo:2020kzh,snowmass2020,Aguilera-Verdugo:2020kzc,Aguilera-Verdugo:2020nrp,Ramirez-Uribe:2020hes,Sborlini:2021owe,Bobadilla:2021pvr,TorresBobadilla:2021ivx,Aguilera-Verdugo:2021nrn,Ramirez-Uribe:2021ubp}. This progress was based on its most remarkable property, the existence of a manifestly causal representation, which was conjectured for the first time in Ref.~\cite{Verdugo:2020kzh}.

In the direct LTD representation noncausal singularities cancel explicitly among all dual terms, nevertheless they lead to considerable numerical instabilities. Regarding causal LTD representation scenario, noncausal singularities are absent and lead to more stable integrands~\cite{Aguilera-Verdugo:2020kzc,Ramirez-Uribe:2020hes}.
Thereby, in this work we combine the most recent developments in LTD with the study of quantum algorithms in perturbative quantum field theory. 

\section{Loop-Tree Duality}
Loop integrals and scattering amplitudes, with $P$ external legs, in the Feynman representation are denoted as integrals in the Minkowski space of $L$ loop momenta
\beq
{\cal A}_F^{(L)} = \int_{\ell_1 \ldots \ell_L} {\cal N} (\{\ell_s\}_L, \{p_j\}_P) \prod_{i=1}^n G_F(q_i)~.
\label{eq:AF}
\eeq
Eq.~(\ref{eq:AF}) is written in accordance with Ref.~\cite{Ramirez-Uribe:2021ubp}. The integration measure in dimensional regularization\cite{Bollini:1972ui,tHooft:1972tcz} is given by $\int_{\ell_s} = -\imath \mu^{4-d} \int d^d \ell_s/(2\pi)^d$, where $d$ is the number of space-time dimensions, and $\mu$ is an arbitrary energy scale. 
Feynman propagators are rewritten in such a way that the poles are shown explicitly
\beq
G_F(q_i) = \frac{1}{(q_{i,0}-\qon{i})(q_{i,0}+\qon{i})}~,
\label{eq:propagator}
\eeq
where $\qon{i} = \sqrt{\qb_i^2+m_i^2-\ii}$, with $\qb_i$ the spacial components of $q_i$ and $m_i$ the mass of the propagating particle. 
From Eq.~(\ref{eq:propagator}) it follows that the integrand in \Eq{eq:AF} becomes singular when the energy component $q_{i,0}$ takes one of the two values $\pm\qon{i}$, this action corresponds to set on shell the Feynman propagator with positive or negative energy.

At one loop the direct LTD representation of \Eq{eq:AF} is calculated by straightforward applying the Cauchy's residue theorem; in a multiloop scenario it is obtained by the evaluation of nested residues~\cite{Verdugo:2020kzh,Ramirez-Uribe:2020hes}.
The selection of the loop momenta component to be integrated is over the energy component given the advantage to work in the integration domain of the Euclidean loop three-momenta space instead of a Minkowski space.

To obtain the causal LTD representation we sum over all the nested residues, after this, the noncausal contributions are explicitly cancelled and the loop integral in \Eq{eq:AF} takes the following form
\beq
{\cal A}_D^{(L)} = \int_{\vec \ell_1 \ldots \vec \ell_L} 
\frac{1}{x_n} \sum_{\sigma  \in \Sigma} \frac{{\cal N}_{\sigma(i_1, \ldots, i_{n-L})}}{\lambda_{\sigma(i_1)} \cdots \lambda_{\sigma(i_{n-L})}}
+ (\lambda_p^+ \leftrightarrow \lambda_p^-)~,
\label{eq:AD}
\eeq
with $x_n = \prod_n 2\qon{i}$.
The Feynman propagators from \Eq{eq:AF} are substituted in \Eq{eq:AD} by causal propagators $1/\lambda_p^\pm$, with
\beq
\lambda_p^\pm = \sum_{i\in p} \qon{i} \pm k_{p,0}~,
\eeq
where $p$ is a set of the on-shell energies, and $k_{p,0}$ is a linear combination of the external momenta energy components. 
Given the sign of $k_{p,0}$, either $\lambda_p^-$ or $\lambda_p^+$ becomes singular after all the propagators in $p$ are set on shell. 
The combinations of entangled causal propagators are collected in the set $\Sigma$, which represent causal thresholds that can occur simultaneously. 

Before going forward let us recall the concept of eloop~\cite{TorresBobadilla:2021ivx,Sborlini:2021owe}, a loop diagram made of edges. We define an edge as the union of an arbitrary number of propagators connecting two interaction vertices.
The selected multiloop topologies to work with are considered in terms of eloops given that in the causality context the only possible causal singular configurations are those in which the momentum flow of all the propagators in an edge are aligned in the same direction. 

\begin{figure}[t]
\center{
\includegraphics[width=280px]{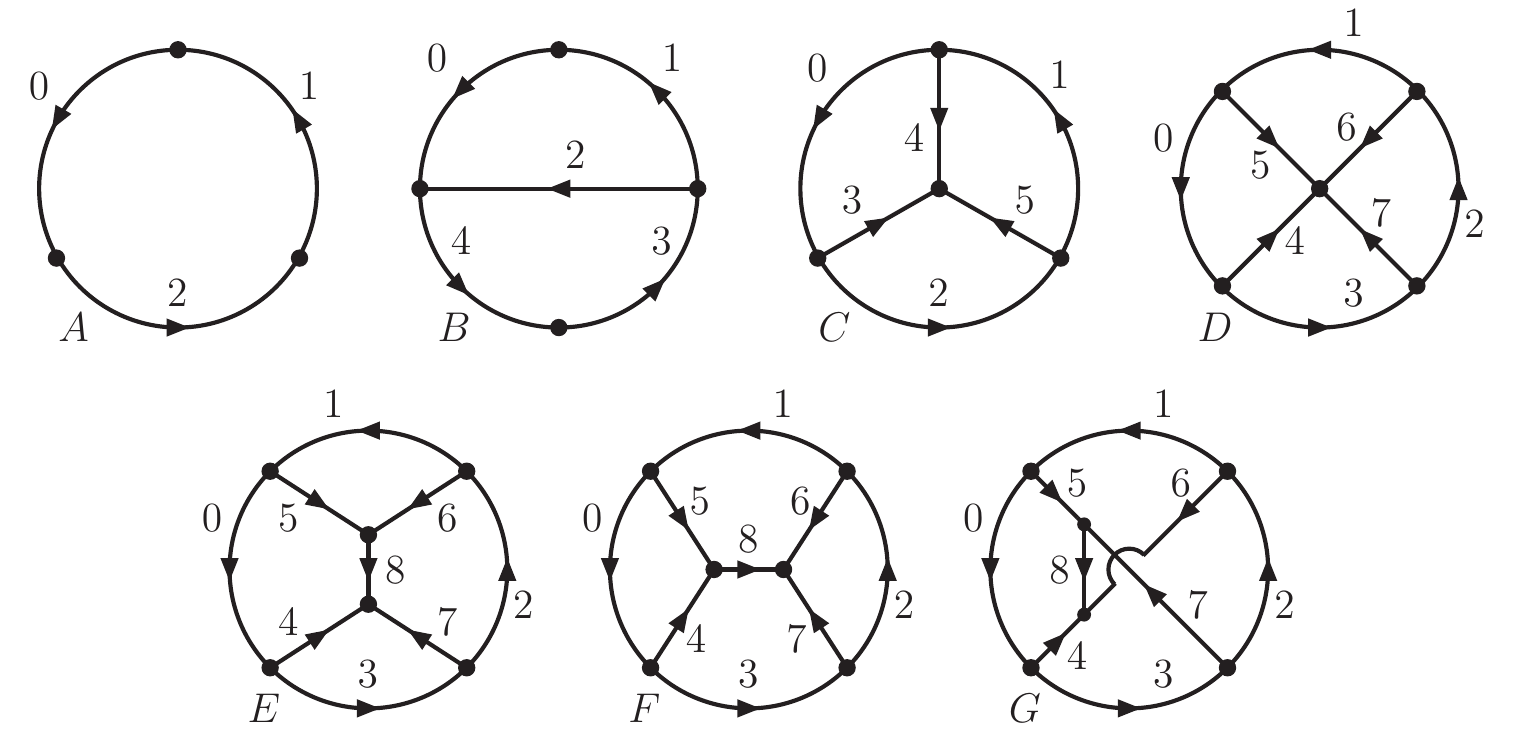}}
\caption{Selected multiloop topologies up to four eloops. The direction of the arrows corresponds to the $\ket{1}$ states. External momenta are not shown. 
\label{fig:Diagram_QC}}
\end{figure}

\section{Modified Grover's quantum algorithm}
Feynman loop integrals can be seen from a quantum computing point of view given the fact that a Feynman propagator has only two possible on-shell states which can be encoded in a qubit, $\ket{1}$ representing states with a specific initial momentum flow configuration and $\ket{0}$ for those with inverse flow orientation. 

The identification of the causal singular configurations can be understood as a query over unstructured datasets~\cite{Boyer:1996zf}. In this work we explored the application of Grover's quantum algorithm~\cite{Grover:1997fa} through the selected multiloop diagrams with the initial configurations shown in Figure~\ref{fig:Diagram_QC}.  

\subsection{Grover's quantum algorithm}
Grover's quantum algorithm is based in three concepts: uniform superposition of all the possible states, an oracle operator to identify the elements searched and a diffusion operator to amplify the probability of these elements.

\begin{enumerate}
\item The uniform superposition of all the $N=2^n$ states is denoted by
$\ket{q}= \frac{1}{\sqrt{N}} \sum_{x=0}^{N-1} \ket{x}$ and also can be written as
\beq
\ket{q} =  \cos \theta \, \ket{q_\perp} + \sin\theta \, \ket{w}~,
\eeq
where $\ket{w}$ and $\ket{q_\perp}$ are the uniform superposition of the winning (causal) and orthogonal (noncausal) states respectively. A crucial element in the algorithm is the mixing angle between these two states, $\theta = \arcsin \sqrt{r/N}$, with $r$ the number of causal states. 

\item The oracle operator, $U_w = \id - 2\ket{w} \bra{w}$, flips the state $\ket{x}$ if $x\in w$, $U_w \ket{x} = - \ket{x}$; and if $x\notin w$, $U_w \ket{x} = \ket{x}$.

\item The diffusion operator, $U_q = 2 \ket{q} \bra{q} - \id$, amplifies the probability of the causal singular configurations by performing a reflection around the initial state $\ket{q}$. 
\end{enumerate}
The iteration of ii) and iii) $t$ times gives
\beq\label{eq:iteration}
(U_q U_w )^t \ket{q} = \cos \theta_t \, \ket{q_\perp } +  \sin \theta_t \, \ket{w}~,
\eeq
with $\theta_t = (2t +1) \, \theta$. 
The mixing angle is critical to define a proper number of iterations. In order to obtain orthogonal state probabilities much smaller than causal state probabilities, $\theta_t$ has to be in accordance with
\beq
\frac{\cos^2 \theta_t}{N-r} \ll \frac{\sin^2 \theta_t}{r}~.
\label{eq:componentprob}
\eeq
Based on Eq.~(\ref{eq:componentprob}), $\theta\leq \pi /6 ~(r/N \leq 1/4)$ allows a good performance on the amplitude amplification provided by the standard Grover's algorithm, on the opposite case the algorithm does not perform well.

Given the selected topologies (see Fig.~\ref{fig:Diagram_QC}) we know for classical computation~\cite{Aguilera-Verdugo:2020kzc,Ramirez-Uribe:2020hes} that the number of causal states is greater than $N/4$.
Nevertheless, there are two adjustments that can be implemented to reduce the number of causal states. The first one previously discuss in Ref.~\cite{Nielsen2000}, is to introduce an ancillary quibit in the $\ket{q}$ register to increase the total number of states without introducing additional solutions. 
The second one is to take advantage of the causal configuration features; given one causal solution, the mirror configuration with all the momentum flows reversed, is also a causal solution~\cite{Ramirez-Uribe:2021ubp}.

The proposal of the modified Grover's quantum algorithm needs three registers and one extra qubit used as a marker in the oracle. The register encoding the $n$ edges is given by $q_i$. The second register is $\ket{c}$ which stores binary clauses, these clauses are labeled as $c_{ij}$ or $\bar{c}_{ij}$ and allow to compare the orientation of two adjacent edges, 
\beq 
c_{ij} \equiv (q_i = q_j),\quad \bar c_{ij} \equiv (q_i \ne q_j)~,
\eeq
with $i,j \in\{0, \ldots, n-1\}$.
The $\ket{a}$ register stands for loop clauses. This register is applied with a multi-Toffoli gate (comparing qubits from $\ket{c}$), used to corroborate if all subloop configurations generate a cyclic circuit. 
The overall scheme of the algorithm is described below:
\begin{enumerate}
\item We have as a first step to initialize all the registers involved in the algorithm. The registers $\ket{a}$, $\ket{c}$ are set to $\ket{0}$ and the qubits standing for the edges are set in a uniform superposition through the Hadamard gate, $\ket{q} = H^{\otimes n} \ket{0}$. The remaining qubit, the Grover's marker is set to the Bell state, $\ket{out_0} = \left(\ket{0} - \ket{1}\right)/\sqrt{2}$.

\item The states of adjacent edges are compared and the validation is stored in the register $\ket{c}$. To implement $\bar c_{ij}$ we need two CNOT gates which perform between $q_i$, $q_j$ and a qubit in the $\ket{c}$ register. For the binary clause $c_{ij}$, an extra XNOT gate is needed to operate on the corresponding qubit in $\ket{c}$. 

\item A function encoding all the causal restrictions is defined, $f(a,q)$. If the causal state conditions are satisfied then $f(a,q) = 1$, if not $f(a,q) = 0$. In addition to the causal restrictions, this function may consider further constraints related to the adjustment in the number of causal states. After defining all the winning conditions, the oracle is implemented as follows 
\bea
U_w \ket{q} \ket{c} \ket{a} \ket{out_0} = \ket{q} \ket{c} \ket{a} \ket{out_0 \otimes f(a,q)}~,
\eea
with
\beq
\ket{out_0 \otimes f(a,q)}=\left\lbrace
	\begin{aligned}
	-\ket{out_0},& \quad {\rm if}~ q \in w ~, \\
	\ket{out_0},& \quad {\rm if}~ q \not\in w~. 
	\end{aligned}
\right.
\eeq
At this point, the causal states are marked and the operations of the oracle are applied in opposite order.

\item Prior to measuring, the diffuser operator is applied to $\ket{q}$. The definition of this operator is taken from IBM documentation provided in the website of IBM Qiskit (\texttt{https://qiskit.org/}). 
\end{enumerate}

\subsection{One eloop}
The first topology to work with is the one-eloop topology consisting of three vertices connected with three edges along one loop. The binary clauses needed are $2$ and there is only one Boolean condition that has to be validated
\beq
a_0(\{c_{ij}\}) \equiv \neg \left( c_{01} \wedge c_{12} \right)~.
\eeq
The qubit $a_0$ is set to one if all the edges are not oriented in the same direction.
This condition is implemented by imposing a multicontrolled Toffoli gate followed by a XNOT gate.

We know that in this case this condition is fulfilled for $6$ states, therefore, the initial Grover's angle tends to $\pi/2$.
In order to achieve the suppression of the orthogonal states, we introduce one ancillary qubit, $q_3$, and select one of the states of one of the qubits representing one of the edges. 
The required Boolean marker is given by 
\beq
f^{(1)}(a,q) = a_0 \wedge q_0 \wedge q_{3}~,
\eeq
which is also implemented through a multicontrolled Toffoli gate. 

\begin{figure}[htb]
\centerline{
\includegraphics[width=160px]{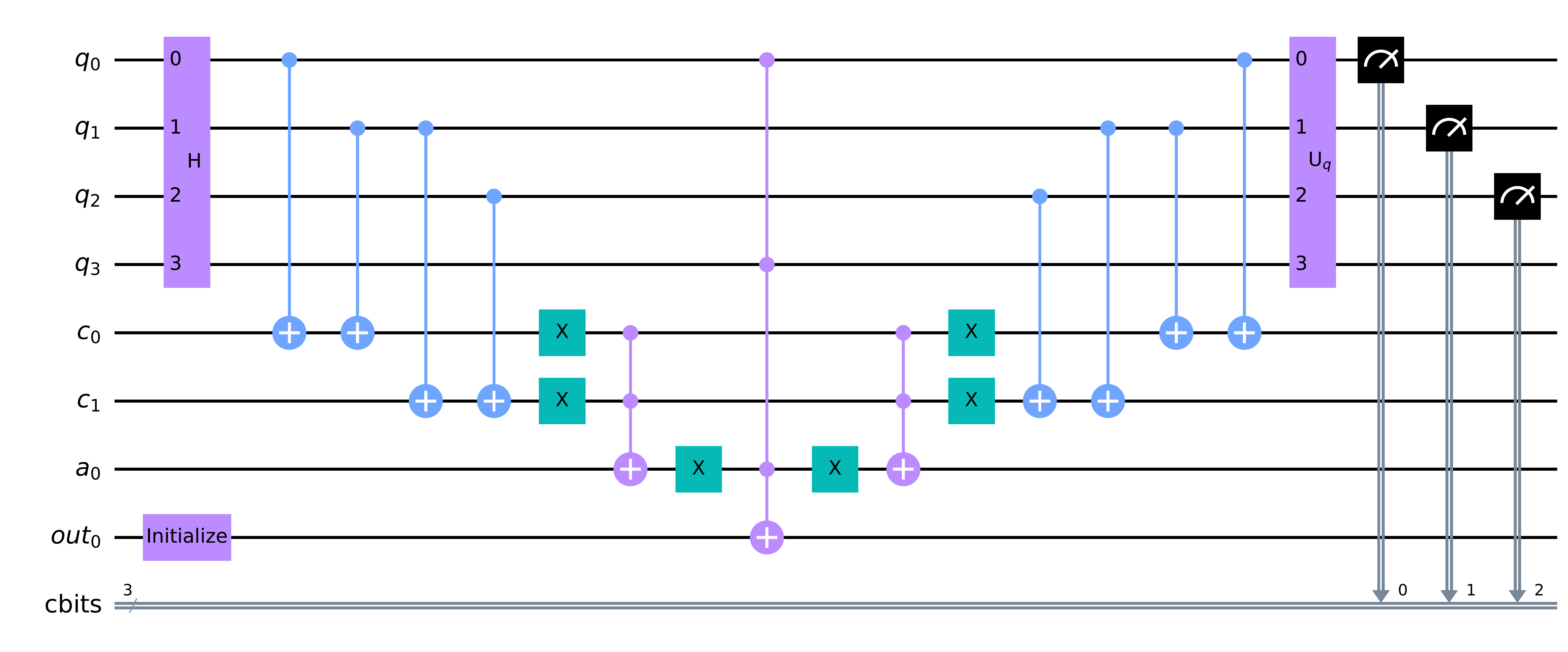}
\includegraphics[width=180px]{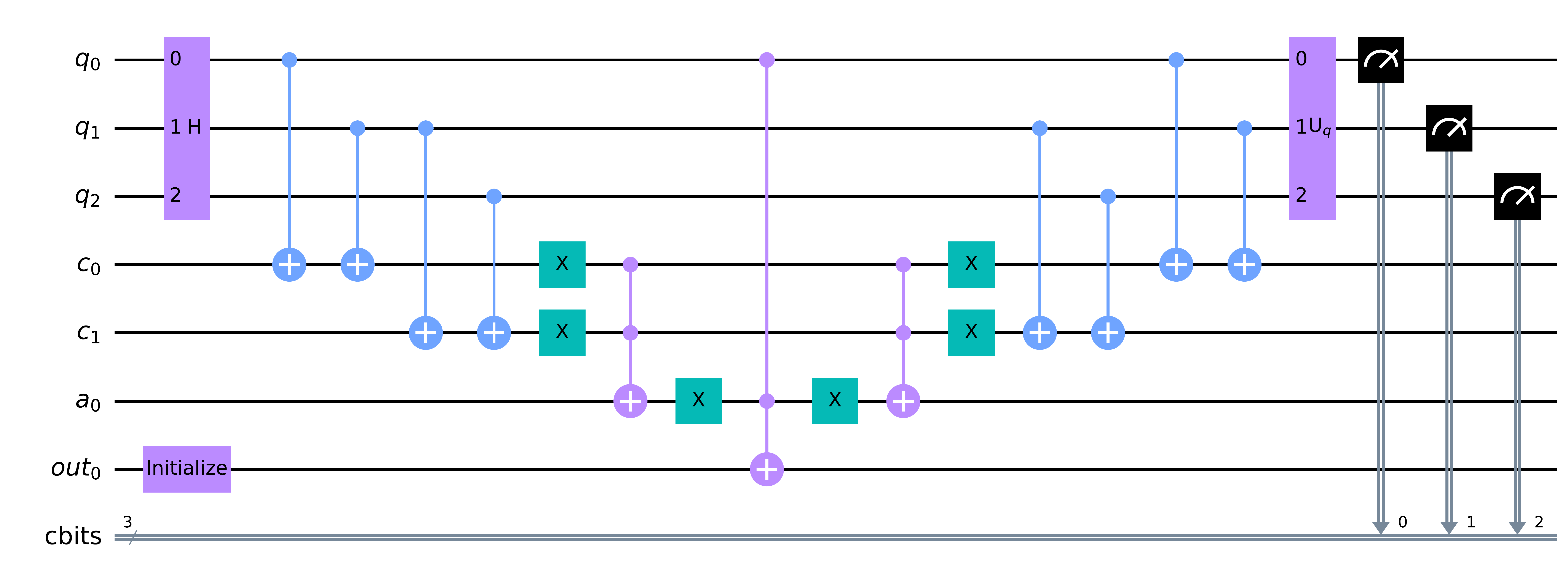}}
\caption{Quantum circuits, with (left) and without (right) an ancillary qubit, used to bootstrap the causal configuration of a three-vertex one-eloop Feynman diagram. \label{fig:qc3edge}}
\end{figure}

For an arbitrary number of edges the Boolean conditions are set as
\bea
a_0(\{c_{ij}\}) &\equiv & \neg \left( c_{01} \wedge c_{12} \wedge \ldots \wedge c_{n-2,n-1} \right)~, \nonumber \\
f^{(1)}(a,q) &=& a_0 \wedge q_0 \wedge q_{n}~.
\eea

\begin{figure}[htb]
\center{
\includegraphics[width=195px]{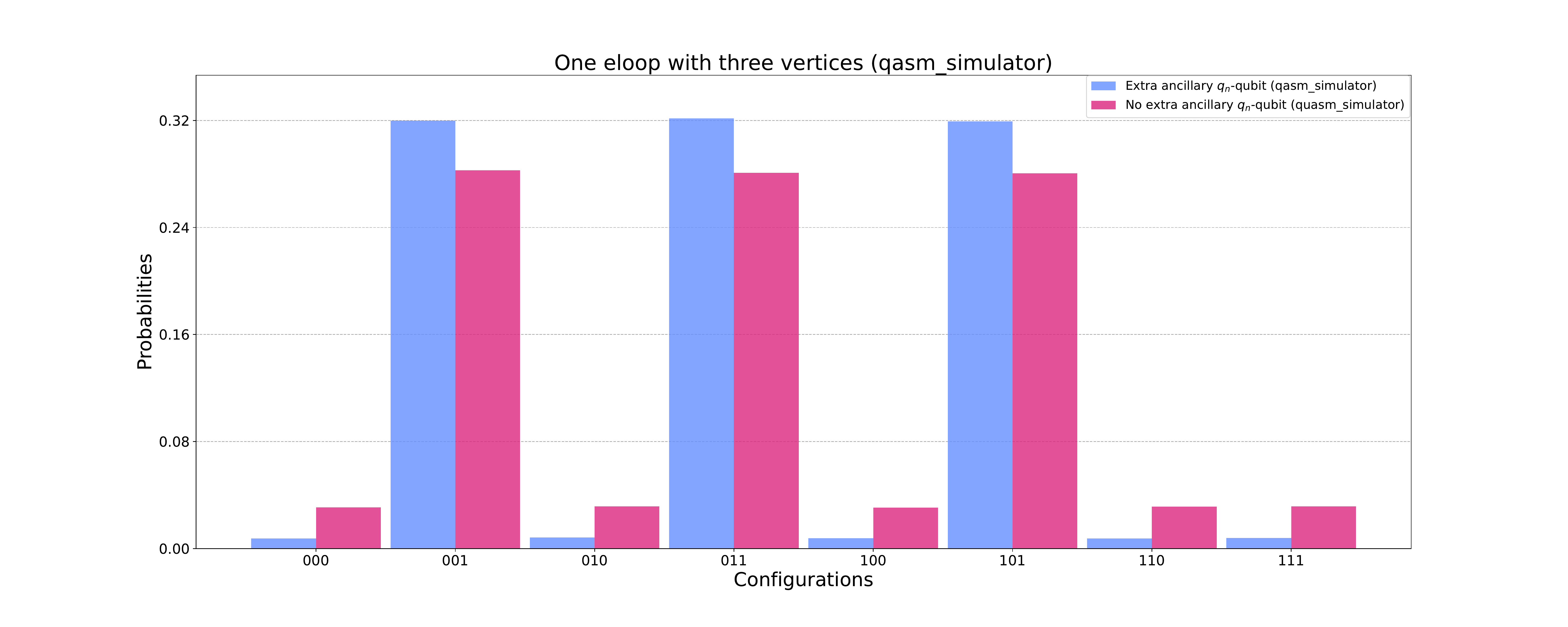}
\raisebox{10pt}{
\includegraphics[width=155px]{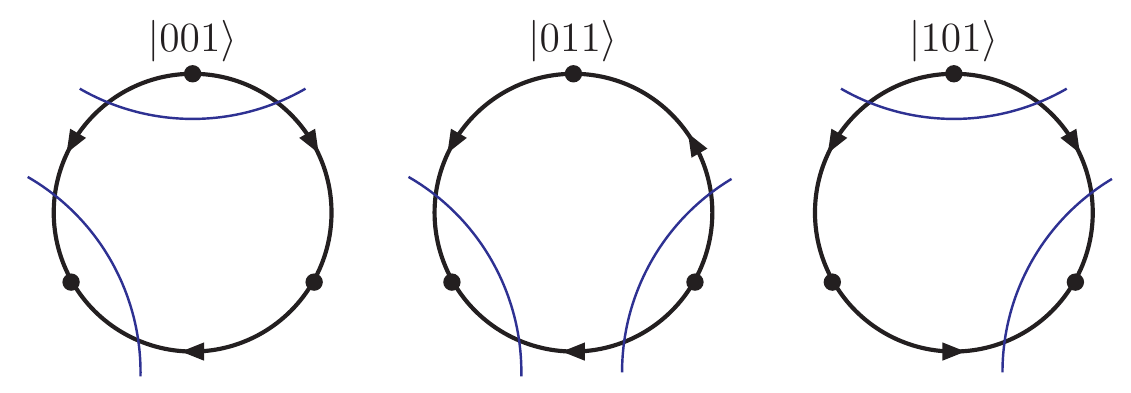}}
}
\caption{From left to right, probability distribution with (blue) and without (purple) an ancillary qubit for a three-vertex one-eloop topology and its translation in terms of causal thresholds.
\label{fig:prob3edge}}
\end{figure}

The corresponding quantum circuits with and without ancillary quibit are depicted in Fig.~\ref{fig:qc3edge}. Together with the qubits representing the edges, the ancillary qubit is set in superposition but is not measured given the irrelevance of the information. The output of the given algorithm and the directed configurations interpreted in terms of causal thresholds are illustrated in  Fig.~\ref{fig:prob3edge}.

An alternative to generate the causal thresholds is through the output of the quantum algorithm, taking into account all feasible cuts with aligned edges that are compatible with each other. This information can be translated directly into the LTD causal representation in \Eq{eq:AD}; the on-shell energies $\qon{i}$ that contribute to a specific causal propagator $( \lambda_p^\pm )$, are those related through the same threshold.

\subsection{Two eloops}
There is analyzed the first nontrivial configuration at two eloops, the one that involves five edges (two of the sets containing two edges).
The diagram is depicted in Fig.~\ref{fig:Diagram_QC}B and is composed by three subloops, therefore the conbination of binary clauses required are
\bea
&& a_0 = \neg \left( c_{01} \wedge c_{13}\wedge c_{34}\right)~, \quad a_1 = \neg \left( c_{01} \wedge \bar c_{12} \right)~, \quad a_2 = \neg \left( c_{23} \wedge c_{34} \right)~. \nn \\
\label{eq:astwoeloops}
\eea
From a classical computation~\cite{Aguilera-Verdugo:2020kzc} we have that the proportion between causal solutions and
the total of states is $18/32 \sim 1/2$, therefore, the use of an ancillary qubit is not needed. The strategy followed is to fix the state associated to $q_2$, giving as a Boolean condition
\beq
f^{(2)}(a,q) = (a_0 \wedge a_1 \wedge a_2) \wedge q_2~.
\eeq

\begin{figure}[htb]
\center{
\raisebox{14pt}{
\includegraphics[width=231px, height=100px]{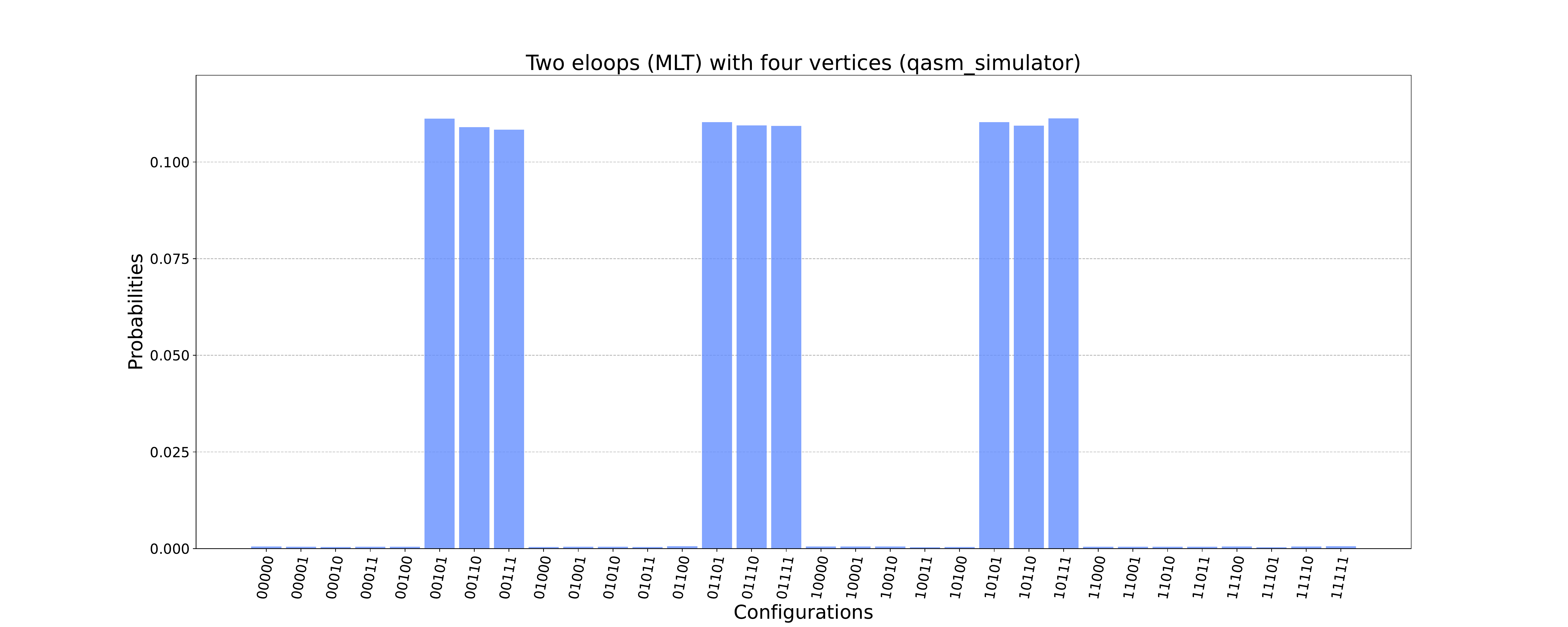}}
\includegraphics[width=119px]{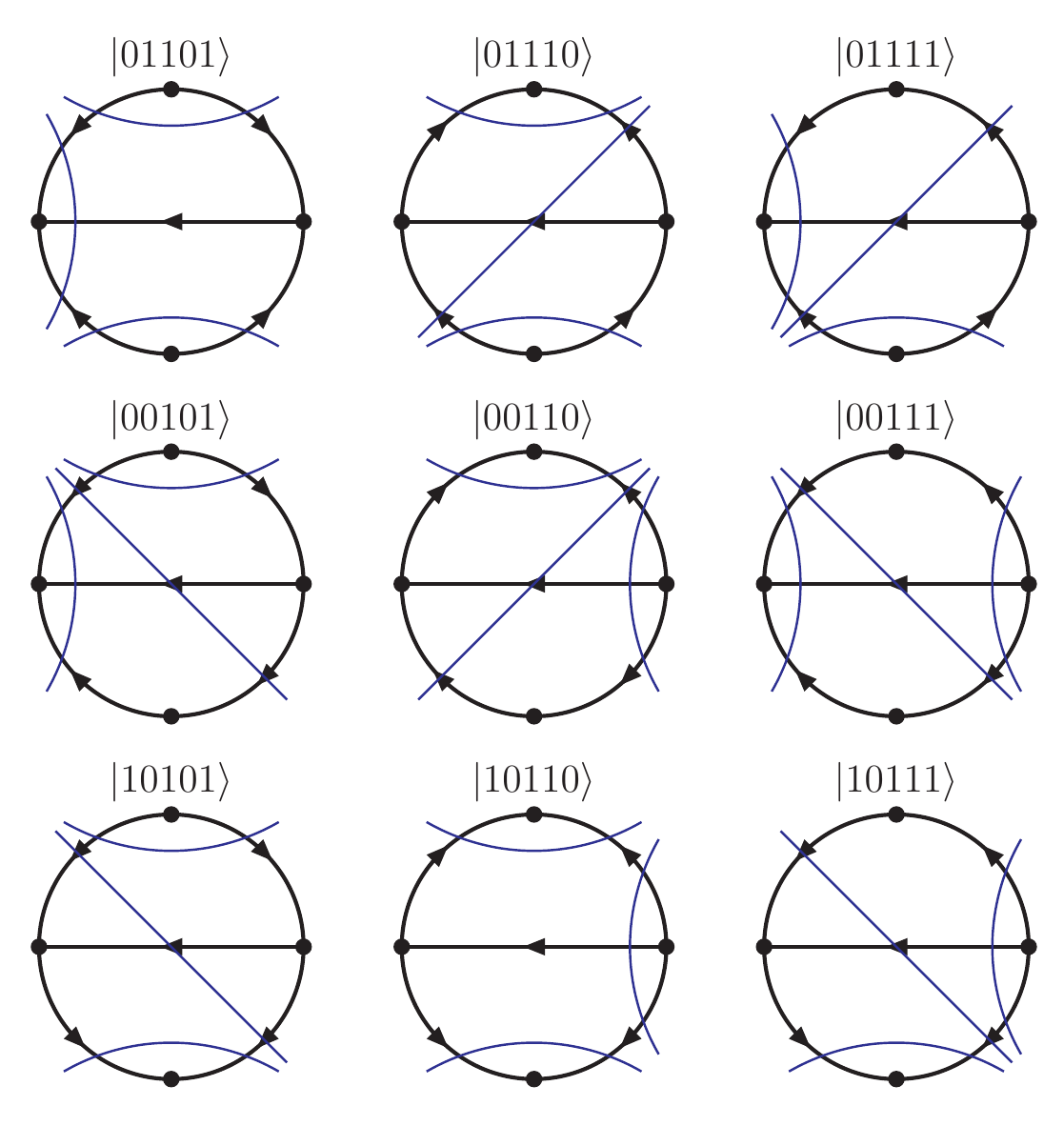}
}
\caption{Probability distribution output of the quantum circuit (left) and the associated entangled causal thresholds (right) for a two-eloops (MLT) topology.
\label{fig:mlt5}}
\end{figure}

The output in the IBM's Qiskit simulator and the causal thresholds interpretation are shown in Fig.\ref{fig:mlt5}. The number of states selected is $9$, corresponding to $18$ causal states when the mirror configurations are considered.

\subsection{Three eloops}

The N$^2$MLT multiloop topology appears for the first time at three loops, characterized by four vertices connected through six sets of edges. The algorithm applied for the multiloop topology shown in Fig.~\ref{fig:Diagram_QC}C, with one edge by set, requires to test the following loop clauses
\bea
&& a_0 = \neg \left( c_{01} \wedge c_{12}\right)~, \quad \quad a_1 = \neg \left( \bar c_{04} \wedge \bar c_{34} \right)~,\nn \\
&& a_2 = \neg \left( \bar c_{15} \wedge \bar c_{45} \right)~, \quad \quad a_3 = \neg \left( \bar c_{23} \wedge \bar c_{35} \right)~.
\label{eq:a3eloops}
\eea

\begin{figure}[t]
\center{
\raisebox{12pt}{
\includegraphics[width=195px, height=90px]{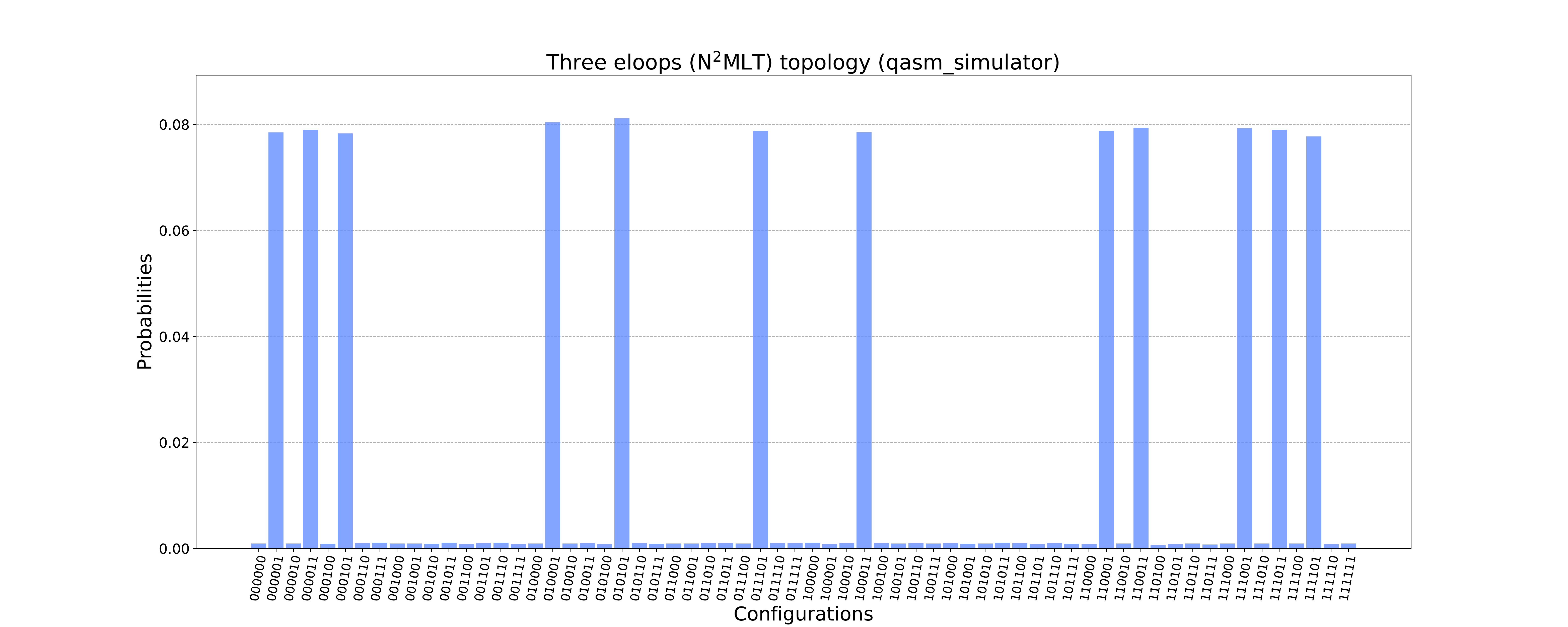}}
\includegraphics[width=155px]{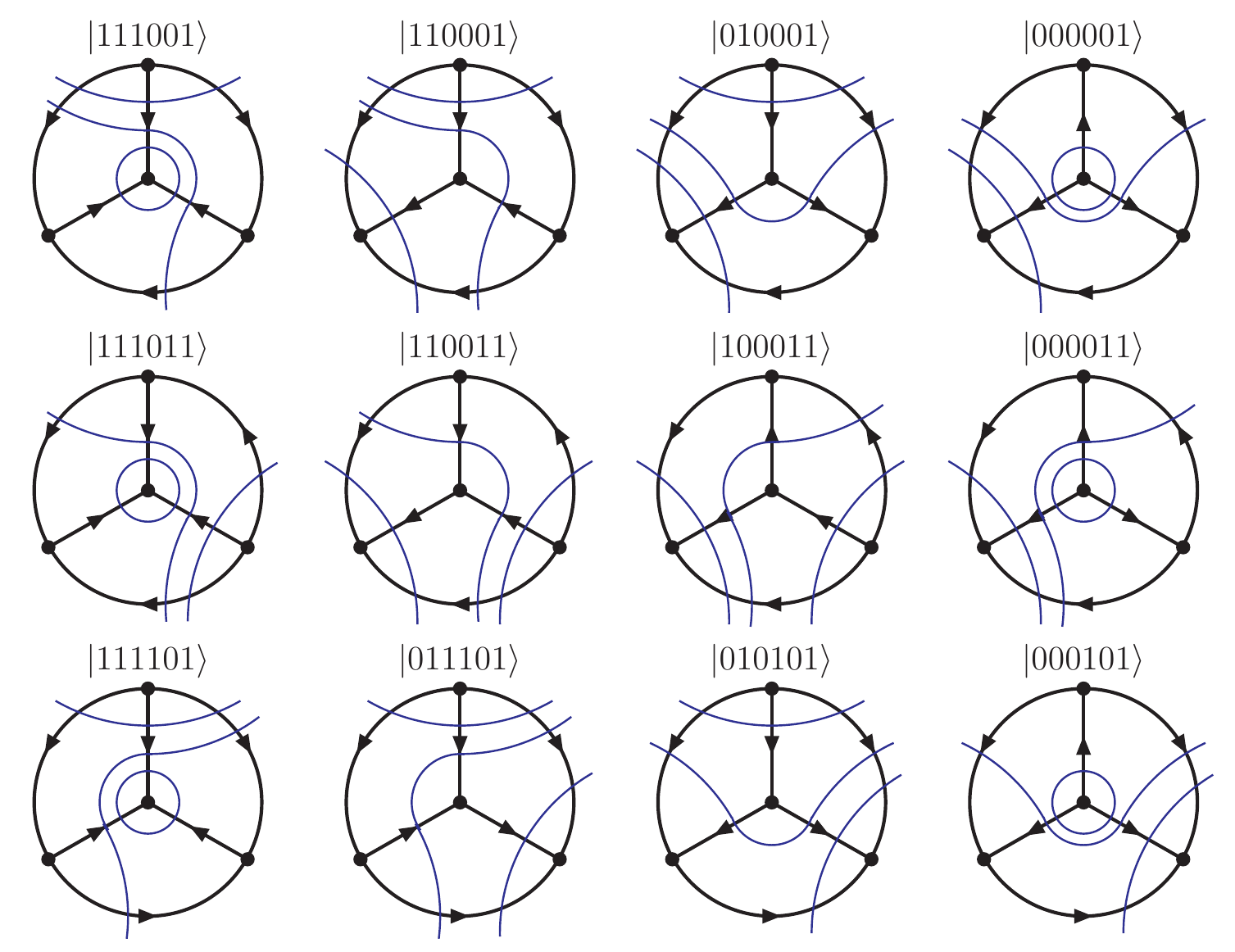}}
\caption{Probability distribution output of the quantum circuit (left) and the associated entangled causal thresholds (right) for the N$^2$MLT topology with one edge by set.}
\label{fig:n2mlt}
\end{figure}

The final Boolean condition is 
\beq
f^{(3)}(a,q) = (a_0 \wedge \ldots \wedge a_3) \wedge q_0~.
\eeq
The probability distribution and the associated causal thresholds are shown in Fig.~\ref{fig:n2mlt}. The total number of causal configurations is $24$ out of $64$ total configurations. For three-eloop configurations with several edges in each set, an extra binary clause and testing loop clauses involving four edges may be needed.

\subsection{Four eloops}

The study of the topologies at four loops is done through the multiloop N$^3$MLT and $t$, $s$ and $u$ channels depicted from Fig.~\ref{fig:Diagram_QC}D to Fig.~\ref{fig:Diagram_QC}G respectively.
The N$^3$MLT multiloop topology is characterized by 8 sets of edges connected through $5$ vertices. 
For a single edge by set the loop clauses are
\bea
&& a_0^{(4)} = \neg \left( c_{01} \wedge c_{12}\wedge c_{23} \right)~, \nn \\
&& a_1^{(4)} = \neg \left( \bar c_{05} \wedge \bar c_{45} \right)~, \quad \quad \quad a_2^{(4)} = \neg \left( \bar c_{16} \wedge \bar c_{56} \right)~, \nn \\
&& a_3^{(4)} = \neg \left( \bar c_{27} \wedge \bar c_{67} \right)~, \quad \quad \quad a_4^{(4)} = \neg \left( \bar c_{34} \wedge \bar c_{47} \right)~. \label{eq:4loopclauses}
\eea

and the Boolean test function
\beq
f^{(4)}(a,q) = (a_0^{(4)} \wedge \ldots \wedge a_4^{(4)}) \wedge q_0~.
\eeq
Some of the loop clauses in \Eq{eq:4loopclauses} are common to the $t$, $s$ and $u$ channels. The channel specific loop clauses needed are 
\bea
&& a_1^{(t)} = \neg \left( \bar c_{05} \wedge \bar c_{45} \wedge \bar c_{48} \right), \quad \quad  a_3^{(t)} = \neg \left( \bar c_{27} \wedge \bar c_{67}  \wedge \bar c_{78} \right)~, \nn \\
&& a_2^{(s)} = \neg \left( \bar c_{16} \wedge \bar c_{56} \wedge \bar c_{68} \right)~, \quad \quad a_4^{(s)} = \neg \left( \bar c_{34} \wedge \bar c_{47}  \wedge \bar c_{78} \right)~, \nn \\
&& a_3^{(u)} = \neg \left( \bar c_{27} \wedge c_{78} \wedge \bar c_{68} \right)~,  \quad \quad
a_4^{(u)} = \neg \left( \bar c_{34} \wedge \bar c_{48}  \wedge c_{78} \right)~, \nn \\
&& a_5^{(u)} = \neg \left( c_{01} \wedge \bar c_{16}  \wedge \bar c_{46} \right)~, \quad \quad
a_6^{(u)} = \neg \left( c_{12} \wedge \bar c_{27}  \wedge \bar c_{57} \right)~, \nn \\
&& a_7^{(u)} = \neg \left( c_{23} \wedge \bar c_{34}  \wedge \bar c_{46} \right)~, \quad \quad
a_8^{(u)} = \neg \left( c_{03} \wedge \bar c_{05}  \wedge \bar c_{57} \right)~.
\eea

\begin{figure}[t]
\center{
\includegraphics[width=160px]{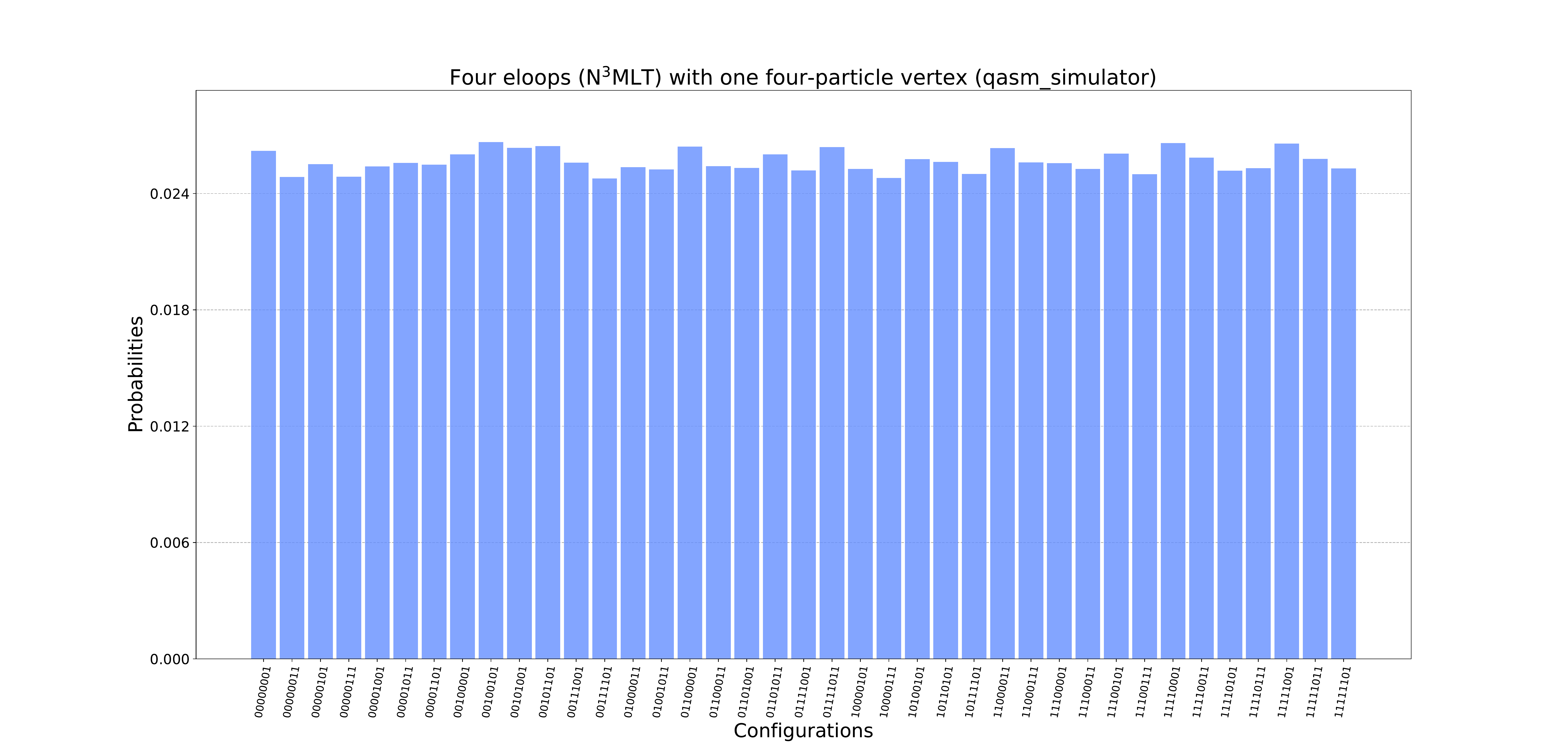}
\raisebox{13pt}{
\includegraphics[width=190px]{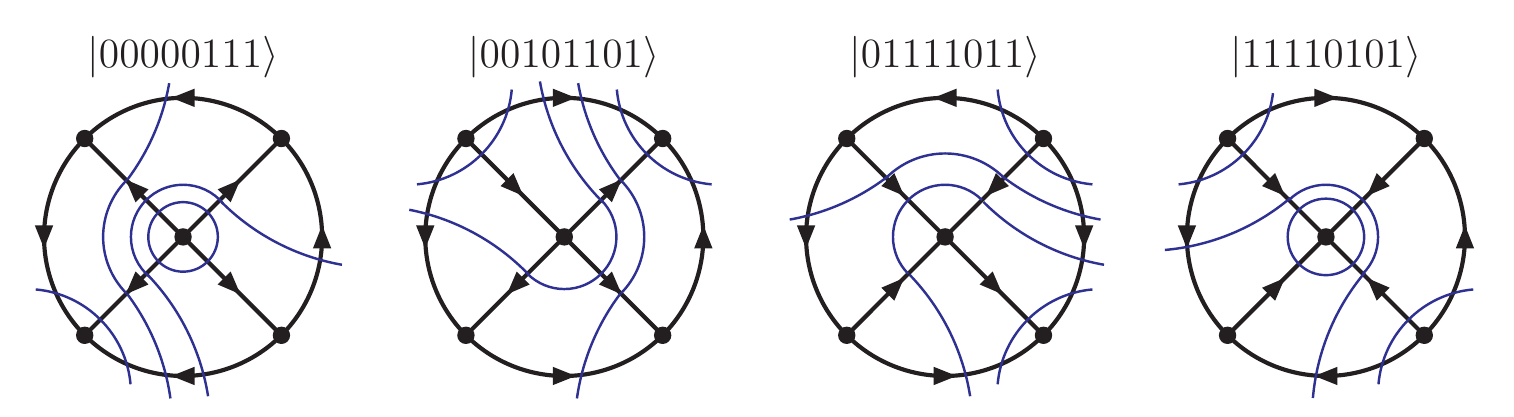}}
}
\caption{Probability of causal configurations (left) and representative entangled causal thresholds (right) from the quantum algorithm applied to the N$^3$MLT topology with one edge by set.} 
\label{fig:prob_u}
\end{figure}

The specific Boolean conditions for each of the $t$, $s$ and $u$ channels are
\bea
&& f^{(4,t)}(a,q) = \left(a_0^{(4)} \wedge a_1^{(t)} \wedge a_2^{(4)} \wedge a_3^{(t)} \wedge a_4^{(4)} \right) \wedge q_0~, \nn \\
&& f^{(4,s)}(a,q) = \left(a_0^{(4)} \wedge a_1^{(4)} \wedge a_2^{(s)} \wedge a_3^{(4)} \wedge a_4^{(s)} \right) \wedge q_0~, \nn \\
&& f^{(4,u)}(a,q) = \left(a_0^{(4)} \wedge a_1^{(t)} \wedge a_2^{(s)} \wedge a_3^{(u)} \wedge \ldots \wedge a_8^{(u)} \right) \wedge q_0~.
\eea
The number of qubits that the algorithm requires for each configuration are $25$, $28$, $28$ and $33$ respectively. The multiloop N$^3$MLT probability of the causal states and representative entangled causal thresholds are shown in Fig.~\ref{fig:prob_u}. The $t$ and $s$ channels are also well supported by the capacity of the IBM's Qiskit simulator. For the $u$ channel the number of qubits needed exceeds Qiskit capacity, in this case the algorithm was implemented within QUTE Testbed framework~\cite{alonso_raul_2021_5561050} which supports up to 38 qubits.

\section{Conclusions}
An application of a quantum algorithm to Feynman loop integrals has been described in detail.
There has been used a modified Grover's quantum algorithm to the identification of the causal singular configurations of selected multiloop topologies up to four loops. 
The proposed algorithm, through the IBM Qiskit and QUTE Testbed quantum simulators, efficiently identifies all causal states for all the multiloop configurations considered. 

The performance of this proposal is of great relevance to the LTD formalism, as it helps to bootstrap the causal representation of mutiloop scattering amplitudes in the loop-tree duality.

\section*{Acknowledgments}
I would like to thank R. Hern\'andez-Pinto, G. Rodrigo and G. Sborlini for the guidance through the development of this work. Also, I am very grateful to CTIC for granting me access to their simulator Quantum Testbed (QUTE) and IBMQ.
Support for this work has been received in part by MCIN/AEI/10.13039/501100011033, Grant No. PID2020-114473GB-I00, COST Action CA16201 PARTICLEFACE, Project No. A1- S-33202 (Ciencia B\'asica), Consejo Nacional de Ciencia y Tecnología and Universidad Aut\'onoma de Sinaloa.

\bibliographystyle{JHEP}

\providecommand{\href}[2]{#2}\begingroup\raggedright\endgroup

\end{document}